\documentclass{mem}
\usepackage{natbib}\usepackage{txfonts}\usepackage{balance}
\usepackage{graphicx}
\usepackage{epstopdf}
\usepackage[a4paper,breaklinks,dvipdfm]{hyperref}
\idline{75}{282}
\begin{document}
\def\teff{$T\rm_{eff }$}
\def\kms{$\mathrm {km s}^{-1}$}

\title{
Cosmic ray escape from supernova remnants
}

   \subtitle{}

\author{
Stefano \,Gabici\inst{1} 
          }

  \offprints{S. Gabici}

\institute{
Astroparticule et Cosmologie (APC), CNRS, Universit\'e Paris 7 Denis Diderot, 
Paris, France.
\email{stefano.gabici@apc.univ-paris7.fr}
}

\authorrunning{Gabici}

\titlerunning{CR escape from SNRs}

\abstract{
Galactic cosmic rays are believed to be accelerated at supernova remnants via diffusive shock acceleration.
Though this mechanism gives fairly robust predictions for the spectrum of particles accelerated at the shock, the spectrum of the cosmic rays which are eventually injected in the interstellar medium is more uncertain and depends on the details of the process of particle escape from the shock. 
Knowing the spectral shape of these escaping particles is of crucial importance in order to assess the validity of the supernova remnant paradigm for cosmic ray origin.
Moreover, after escaping from a supernova remnant, cosmic rays interact with the surrounding ambient gas and produce gamma rays in the vicinity of the remnant itself.
The detection of this radiation can be used as an indirect proof of the fact that the supernova remnant was indeed accelerating cosmic rays in the past. 
}
\maketitle{}


Supernova remnants (SNRs) are believed to accelerate galactic cosmic rays (CRs) up to at least the energy of the knee ($\approx 3$~PeV) via diffusive shock acceleration \citep[e.g.][]{hillas}. Though this is a very popular idea, a conclusive proof of its validity is still missing. The detection of a number of SNRs in GeV and TeV gamma rays \citep[e.g.][]{damiano} is encouraging, since gamma rays are expected to be produced in hadronic interactions between the accelerated CRs and the ambient gas swept up by the shock. However, competing leptonic processes may also explain these observations, and thus the debate on the origin of such emission is still open. Notably, the emission from the SNR RX~J1713.7-3946 seems to be leptonic \citep{fermiRXJ}, the one from Tycho hadronic \citep{fermiTycho}, while for most of the other gamma ray bright SNRs the situation is less clear.

Gamma rays can also be produced by CRs after they escape from SNRs due to hadronic interactions in the ambient gas. Thus, the detection of gamma ray emission from the vicinities of SNRs might provide an indirect evidence for the fact that the nearby SNR is (or was) accelerating CRs \citep{atoyan,gabici07}. Massive molecular clouds have been detected in TeV gamma rays in the vicinity of the SNR W28 \citep{HESSW28}. This might constitute a first example of such situation \citep{gabici10}. Future observations with the Cherenkov Telescope Array will undoubtedly increase the number of detected SNR/cloud associations and clarify the issue \citep{CTA}.

Unfortunately, the way in which CRs escape SNRs is little understood. Qualitatively, CRs with the highest energies are expected to escape first, while CRs of lower and lower energy are expected to escape progressively later in time, as the shock slows down \citep{ptuskinzirakashvili03}, but the details of how this happens are still unclear.
In this paper, the present status of the studies on CR escape from SNRs is reviewed. A review of the aspects related to the gamma~ray radiation produced by runaway CRs in the surrounding of SNRs can be found in Casanova, these proceedings. 

\section*{Qualitative estimates}

A supernova explosion drives an expanding shock in the interstellar medium. As long as the mass of the material ejected in the explosion is much smaller that the mass of the gas swept up by the shock, the shock speed $u_{sh}$ remains roughly constant. This stage of the SNR evolution is called free-expansion phase. If particles are accelerated at the shock via diffusive shock acceleration \citep[e.g.][]{luke}, the acceleration time is: $t_{acc} \approx D/u_{sh}^2$, where $D$ is the particles diffusion coefficient. Assuming Bohm diffusion ($D \propto E/B$, where $E$ is the particle energy and $B$ the magnetic field strength assumed here to be constant in time), as seems appropriate at shocks, and equating the acceleration time to the age of the SNR, $t_{age}$, one gets the maximum energy that a particle can reach at the shock: $E_{max} \propto t_{age}$. Thus, in this phase, the maximum energy increases linearly with time. 

As a consistency check, note that CRs with the highest energy diffuse ahead of the shock one diffusion length $l_d \approx D(E_{max})/u_{sh}$ before returning to it. It is straightforward to show that this length is of the same order of the SNR radius $R_{sh} = u_{sh} t_{age}$.
Thus, by adopting a Hillas-like criterium for CR confinement in the SNR (diffusion length $\lesssim$ SNR size) one can conclude that all the accelerated CRs are effectively confined and particle escape does not occur \footnote{CR escape may indeed occur also during the free expansion phase \citep{ptuskinzirakashvili05} due to the actual, though slow, decrease of the shock velocity with time. However, the number of particles involved in the shock acceleration is quite small in this phase, so in the following we will neglect them.}.

Once the mass of the gas swept up by the shock becomes larger than the mass of the ejecta the SNR enters the adiabatic, or Sedov phase. For an expansion in an uniform medium this phase is characterized by the scalings: $R_{sh} \propto t^{2/5}$ and $u_{sh} \propto t^{-3/5}$. Let us call $E_{max}^*$ the maximum energy of the accelerated particles at the time $t^*$ of the  transition between the free-expansion and the Sedov phases. Such particles are confined within the SNR because their diffusion length is of the same order of magnitude than the SNR radius $l_d(E_{max}^*, t^*) \approx R_{sh}(t^*)$. Once the SNR enters the Sedov phase, the particle diffusion length increases as $l_d \approx D(E^*_{max})/u_{sh} \propto t^{3/5}$, faster than the SNR radius, which scales like $t^{2/5}$. 
Therefore, particles with energy $E^*_{max}$, which were confined during the free expansion phase, now violate the Hillas criterium and can escape the SNR.
This implies that the decrease of the shock velocity reduces the capability of the SNR to confine particles. 

Qualitatively, to see when CRs with different energies leave the SNR, one can equate the particle diffusion length with the SNR radius, to get: 
$E_{max} \propto R_{sh} u_{sh} B \propto t^{-1/5}$.
This very rough estimate has the sole purpose to provide us with a qualitative description of the the process of CR escape from SNRs. 
What can be understood from that is that CRs are released mainly during the Sedov phase, and that they are released gradually, the ones having the highest energy first, and the ones with lower and lower energies at later and later times. 
The decrease with time of the maximum energy of confined particles is quite mild, but this has to be ascribed uniquely to the fact that a diffusion coefficient constant in time has been assumed in the derivation.
If the magnetic field at the shock is amplified by CR streaming instability, the value of the diffusion coefficient is expected to increase as the shock decelerates \citep{ptuskinzirakashvili03}.
As a result, a faster decrease of $E_{max}$ with time may be expected. 

The approach presented here is (deliberately) qualitative and suffers from many limitations. Nevertheless, it gives a fairly reasonable description of the main features and trends which are also derived from detailed models, which will be reviewed in the next Section.

Another important issue related to the process of particle escape from SNRs concerns the shape of the CR spectrum which is eventually released in the interstellar medium.
It is well known from the measurements of the abundances of secondary elements such as Li, Be, and B in the cosmic radiation that the CR confinement time in the Galaxy decreases with energy as $t_{esc} \propto E^{-s}$, with $s \approx 0.3...0.6$.
Thus, to reproduce the shape of the observed CR spectrum $\propto E^{-2.7}$, CR sources have to inject a spectrum with slope close to, but definitely steeper than 2: $N_{inj} \propto E^{-2.7+s} \approx E^{-2.1...2.4}$.
How does this compare with the spectrum that SNRs are expected to inject in the interstellar medium?

To estimate such a spectrum one can proceed as follows \citep[see][for a detailed treatment]{ptuskinzirakashvili05}: let $f_{esc}(t)$ be the fraction of the kinetic energy flux across the SNR shock that is carried away by escaping particles. For a SNR expanding in a homogeneous medium and in the Sedov phase, the energy carried away per unit time by escaping CRs reads: $d{\cal E}/dt = f_{esc} (\rho u_{sh}^3/2) (4 \pi R_{sh}^2) \propto f_{esc} t^{-1}$. Let us now assume that particles of energy $E$ escape at a time $t \propto E^{-\delta}$.
This allow us to write the spectrum of escaping CRs as:
$N_{esc} = (d{\cal E}/dt)/E \times (dt/dE) \propto f_{esc} E^{-2}$.
Thus, if $f_{esc}$ is constant in time, 
the spectrum of the particles released by a SNR in the interstellar medium during the whole Sedov phase is $\propto E^{-2}$, regardless of the value of the parameter $\delta$ which describes the way in which CRs of different energies are released.
Remarkably, such result has beed obtained without specifying the spectral shape of the particles at the shock.

This last point deserves more attention \citep[what follows is a simplified summary of the results derived by][]{ohira}.
Let us now assume, in a slightly different fashion, that a fraction $\eta$ (again, for simplicity, constant in time) of the kinetic energy flowing across the shock per unit time $L_k$ is converted into accelerated particles.
These particles have a spectrum $Q_{CR} \propto E^{-\alpha}$ [\#/time/energy] and after acceleration are advected downstream of the shock.
The source term for escaping particles can be written as: $Q_{esc} \propto E ~ Q_{CR}(E, t) ~ \delta(E-E_{max}(t))$, where we have assumed that only recent acceleration is relevant \citep{ptuskinzirakashvili05}.
In an expanding SNR particles accelerated at earlier times are advected downstream and suffer adiabatic energy losses, and thus their contribution can, as a first approximation, be neglected. 
The energy per unit time in form of CRs that escape the shock is: $F_{esc} = \int Q_{esc} ~ E ~ {\rm d}E \propto Q_{CR}(E_{max}) E_{max}^2$.

First, consider the situation in which the slope of the spectrum of the accelerated CRs $\alpha$ is smaller than 2. 
In this case, the particles with the highest energy carry most of the total energy in form of CRs, and we can write: $\eta L_k \approx Q_{CR}(E_{max}) E_{max}^2 \propto F_{esc}$.
This means that a constant fraction of the shock kinetic energy flux is converted into escaping particle, which, as discussed above, implies that the spectrum of the particles released by the SNR is $N_{esc} \propto E^{-2}$.
Conversely, if the spectrum of the CRs accelerated at the shock is softer than 2, particles with energy $E_0 \sim mc^2$ carry most of the energy, so we have: $\eta L_k \approx Q_{CR}(E_0) E_0^2 \propto Q_{CR}(E_{max}) E_{max}^2 (E_0/E_{max})^{2-\alpha}$. The last term in brackets tells us that $f_{esc} \propto E_{max}^{2-\alpha}$ and therefore in this situation the spectrum of escaping CRs is softer than $\alpha = 2$ and equal to $N_{esc} \propto E^{-\alpha}$.
Thus, in order to be consistent with CR data, not only the spectrum injected in the interstellar medium, but also the spectrum of CRs at the SNR shock has to be steeper than $E^{-2}$. This will be discussed in the next Section.


\section*{Overview of models}

The acceleration of particles at shock waves is believed to be a very efficient mechanism, i.e. a large fraction of the shock kinetic energy flux can be converted into accelerated particles. As a consequence, the pressure term due to the presence of accelerated particles cannot be neglected in the flow equations (as it is done in the test-particle limit), the shock structure is then modified by the presence of CRs, and the whole process of acceleration becomes non-linear \citep[for a review see][]{malkovdrury}.
Extensive efforts have been devoted to the solution of the steady-state equations describing non-linear shock acceleration, and an approximate analytic solution has been found by \cite{malkov}.
Accelerated particles exhibit a concave spectrum, which is harder than $E^{-2}$ (up to $E^{-1.5}$) at the highest energies.
In Malkov's aproach, as well as in several other analytic or semi-analytic approaches, the maximum momentum of accelerated particles, $p_{max}$, is a free parameter, and the particle distribution function vanishes abruptly above that momentum.
Since a flux of particles upward in momentum exists at the shock, this is equivalent to assume that particles accelerated over $p_{max}$ escape the system \citep[see][for an early discussion of this issue]{eichler}.
Within this approximation, it is possible to estimate the amount of energy subtracted from the system by the runaway CRs. 
Due to the hardness of the spectrum, this can amount to a significant fraction of the total bulk energy flowing across the shock \citep[e.g.][]{berezhkoellison,giulia,elena}.

Another possibility to model particle escape is to assume a {\it spatial} escape boundary, instead of the one in energy space described above. 
This is more justified, since the CRs themselves are believed to amplify the magnetic field at the shock and generate the turbulence they need to be scattered off via resonant \citep{cesarsky} or non-resonant \citep{bell04} streaming instability.
Depending on the CR flux at a given position, the level of the magnetic turbulence is expected to decrease with the distance upstream of the shock. This suggests the existence of a spatial boundary beyond which CRs cannot be effectively scattered and freely escape.
\citet{brian} studied the acceleration of CRs at a shock in presence of a spatial escape boundary and assumed the turbulence to be generated at such boundary by the non-resonant streaming instability.
This instability injects short-wavelength turbulence, not very efficient in scattering high energy particles with a large gyroradius, but in its nonlinear evolution is characterized by the appearance of large scale structures like cavities \citep{bell05} which might reduce the mean free path of high energy particles. 
Thus, \citet{brian} made the crude, but not unreasonable, approximation that the particle diffusion coefficient is of the Bohm type in a background magnetic field amplified to the value at which the non-resonant instability is believed to saturate.
Within this approximation, the spatial escape boundary is expected to be roughly located at the position at which the turbulence growth rate equals $u_{adv}/l_p$, where $u_{adv}$ is the velocity at which waves are advected towards the shock and $l_p$ is the distance upstream of the shock.
The non-resonant mode of streaming instability is likely to dominate over the resonant one at large distances from the shock \citep{guy}, and thus its properties should fix the position of the boundary.
 
Though a fully self consistent treatment of the problem is still not available, introducing a spatial escape boundary presents several advantages. 
First, it allows to determine the shape of the cutoff in the CR spectrum and the spectrum of the escaping particles, which are no longer an abrupt step-function feature and a delta function in energy, respectively, as in the case of an energy escape boundary. 
This is very important in order to investigate the radiative signatures related to CR acceleration at shocks, whereas it is of minor relevance for the estimate of the radiation from runaway particles, whose spectrum is found to be quite narrow and peaked around $p_{max}$ \citep{brian,blasiescape3}.
Moreover, it is possible to check, a posteriori, the consistency between the assumed position of the escape boundary and the value of the amplified magnetic field/diffusion coefficient, that depends on the spatial CR intensity \citep{brian}.

Free escape boundaries have been used in Monte Carlo approaches also, both in energy \citep[e.g.][]{ellisoneichler}, and space \citep[e.g.][]{vladimirov}, as well as in time-dependent numerical simulations of shocks \citep[e.g.][]{kang95}, with the value of the maximum energy or the spatial position of the boundary as a free parameter of the models.
A critical discussion on the use of free escape boundaries can be found in \cite{lukeescape}. 

\citet{vladbell} also studied the acceleration of CRs at a plane shock in presence of the non-resonant streaming instability and computed the CR diffusion coefficient in the resulting short-wavelength turbulence. 
A free escape boundary was imposed at a distance $L$ upstream of the shock, with $L$ of the order of the radius $R_s$ of the SNR to be simulated. 
One of the results from this simulation is that the width of the CR distribution upstream of the shock is small, of the order of $\lesssim 0.1 ~ L$, which implies that the CR acceleration at SNRs can be described with satisfactory accuracy  as a one-dimensional phenomenon even in three-dimensional systems such as SNRs.
Moreover, the maximum energy of accelerated CRs was found to be lower than the one that would be obtained by assuming Bohm diffusion in the amplified magnetic field, since the scattering in the short-wavelength turbulence is not so efficient.
In particular, only type Ib/c and IIb supernovae were found able to accelerate CRs up to PeV energies, while for type II-P supernovae, the ones with the highest explosion rate, a maximum energy of only a few hundreds of TeV was obtained.
However, in order to fully understand how particles with the highest energy are confined and accelerated at a shock, and thus how SNR can accelerate particles up to the PeV range and above, a study of the instabilities that involve long-wavelengths (i.e. larger than the gyroradius of the generating CRs) is needed. Promising results in this direction have been recently published  \citep{bykov,clara}.

To conclude this section we discuss, in the light of the detailed models described above, which are the expectations for the total CR spectrum injected by a generic SNR in the interstellar medium.
As discussed in the previous section, in order to be consistent with CR data, CR sources have to inject spectra steeper than $E^{-2}$, with slopes in the range $\approx 2.1...2.4$.
However, also the observed high level of isotropy of CRs has to be taken into account.
For example, an injection spectrum $\propto E^{-2.1}$ corresponds to a CR escape time from the Galaxy that increase quickly with energy $t_{esc} \propto E^{0.6}$.
In this case, the escape time of high energy particles would be so short to result in a strong anisotropy of CR at the Earth, in contrast with observations \citep[e.g.][]{hillas}.
Though the discussion on this is still ongoing and requires further investigation, let us choose as a reference value for what follows a slope of the injection spectrum equal to $\approx 2.4$, which should minimize the problem of anisotropy. 

Such a steep spectrum for the runaway CRs corresponds, roughly, to an equally steep spectrum for the CRs accelerated at the SNR shock (see previous Section).
This seems to be in disagreement with the predictions of non-linear shock acceleration theories, according to which spectra are, at high energies, harder than $E^{-2}$.
A way to solve this problem is the following \citep{alfvendrift}: the spectrum of the accelerated particles is determined by the compression factor of the velocities of the {\it scattering centres}, which can be different from the shock compression factor $r$ (ratio between up- and down-stream {\it gas} velocities), which is often used to determine the CR spectrum.
If the scattering centres move against the fluid upstream and with it downstream (as might be the case for Alfven waves), the compression factor that CRs feel at the shock may be reduced with respect to $r$ and the resulting CR spectrum significantly steepened.

By including this effect, named Alfven drift, in a numerical code describing non-linear CR acceleration at a spherical SNR shock, \citet{ptuskinseo} obtained a good fit to the entire CR spectrum up to an energy of $\sim 3 \times 10^{18} {\rm eV}$, well above the knee.
They considered the contributions from different types of supernovae (Ia, IIP, Ib/c, IIb) and parametrized the value of the magnetic field at the shock in order to mimic the magnetic field amplification, as it is observed from X-ray synchrotron filaments \citep{heinz}.
Bohm diffusion in the amplified field was adopted.
Less optimistic conclusions have been drawn by \citet{blasiescape2}, which employed a (one-dimensional and quasi-steady-state) model for non-linear particle acceleration to compute the spectrum of CRs escaping from individual SNRs, though they considered the magnetic field amplification due to {\it resonant} streaming instability \citep{cesarsky}.
They obtained power law spectra roughly of the order of $E^{-2}$, too hard to explain the observed CR spectrum. When including the effect of Alfven drift, spectra become steeper but do not reach the energy of the knee. Additional (non-resonant?) field amplification might mitigate this problem.

\section*{Gamma rays from runaway CRs}

After escaping the SNR, CRs interact with the ambient gas and produce gamma rays.
The production of such radiation is enhanced if dense gas (e.g. a molecular cloud) is present in the vicinity of the SNR. The CR background in the Galaxy has a steep spectrum $\approx 6 \times 10^{-3} (E/{\rm TeV})^{-0.7} {\rm eV~cm^{-3}}$, and thus an excess in the CR intensity would appear more easily at high energies. For this reason we focus here on the energy domain probed by Cherenkov instruments ($\gtrsim 100$~GeV).

To maintain the CR intensity at the observed level, each SNR has to provide $\approx 10^{50} E_{50}$~erg in form of CRs ($\approx$10\% of the explosion energy). For a CR spectrum with slope 2.4 this corresponds to: $E^2 N_{CR}(E) \approx 2 \times 10^{60} E_{50} (E/{\rm TeV})^{-0.4}$~eV. If CR diffusion proceeds isotropically, particles with a given energy at a given time after escape are  distributed roughly uniformly in a spherical region of radius $R_* \approx \sqrt{6 ~ D ~ t}$ surrounding the SNR. The energy density of runaway CRs exceeds the one in the background if $R_* \lesssim 130 ~ E_{50}^{1/3}$pc. The average diffusion coefficient of TeV CRs in the Galaxy is $\approx10^{29}~{\rm cm^2/s}$ and thus the excess of CRs lasts for $\approx R_*^2/6 D \approx 10^4$~yr.
Therefore, the search for gamma ray emission from molecular clouds located close to SNRs has to be focused on regions of size $100-200$~pc around SNRs not much older than $\approx 10^4$~yrs.

The detection of such radiation can serve as an indirect proof for the fact that SNRs are indeed the sources of CRs. Moreover, the figures given above depend on the {\it local} (close to the SNR) value of the diffusion coefficient, which might differ significantly from the average galactic one (e.g. due to enhanced CR streaming instability in the vicinity of a CR source). Thus, the characteristics of the radiation may also be used to constrain the properties of CR diffusion (see \citealt{gabici09} and Casanova, these proceedings for details).
To date, very few SNR/cloud associations have been detected in TeV gamma rays (e.g. the SNR W28), but future observations with the Cherenkov Telescope Array promise to increase the number of detected objects and provide a test for this scenario.

\begin{acknowledgements}
I am grateful to the organizers of the CRISM conference for their invitation and support. 
Support from the EU is also acknowledged [FP7 - grant agr. n$^o$256464].
\end{acknowledgements}

\bibliographystyle{aa}

\end{document}